\documentstyle[12pt]{article}

\title{The electronic structure of amorphous silica:\\
A numerical study}
\author{Thorsten Koslowski$^1$, 
Walter Kob$^2$ and Katharina Vollmayr$^3$}
\date{$^1$ Institut f\"ur Physikalische Chemie
und Elektrochemie I, \\
Universit\"at Karlsruhe, Kaiserstra\ss e 12, \\
D-76128 Karlsruhe, Germany \\ \ \\
$^2$ Institut f\"ur Physik, Johannes-Gutenberg-Universit\"at, \\
Staudinger Weg 7, D-55099 Mainz, Germany \\ \ \\
$^3$ Institute for Physical Sciences and Technology, \\
University of Maryland, College Park, MD 20742, USA}
\begin{document}
\maketitle
\newpage
\begin{abstract}
We present a computational study of the electronic properties of
amorphous SiO$_2$. The ionic configurations used are the ones generated
by an earlier molecular dynamics simulations in which the system was
cooled with different cooling rates from the liquid state to a glass,
thus giving access to glass-like configurations with different degrees
of disorder [Phys. Rev.  B {\bf 54}, 15808 (1996)].  The electronic
structure is described by a tight-binding Hamiltonian.  We study the
influence of the degree of disorder  on the density of states, the
localization properties, the optical absorption, the nature of defects
within the mobility gap, and on the fluctuations of the Madelung
potential, where the disorder manifests itself most prominently.  The
experimentally observed mismatch between a photoconductivity threshold
of 9~eV and the onset of the optical absorption around 7~eV is
interpreted by the picture of eigenstates localized by potential energy
fluctuations in a mobility gap of $\sim $9~eV and a density of states
that exhibits valence and conduction band tails which are, even in the
absence of defects, deeply located within the former band gap.

\end{abstract}
\newpage
\subsection*{1. Introduction}
\label{sec1}

Apart from being of great importance in chemistry, optics, geology and
industrial applications, silica is a prototype of a strong glass former
and has thus been investigated extensively by experimental, theoretical
and computational techniques.  It is generally believed that a-SiO$_2$
shares its microscopic structure with the low-density crystalline
phases of SiO$_2$ and that bond lengths and bond angles show values
comparable to their crystalline counterparts. The concept of a random
network of corner-sharing tetrahedral SiO$_4$ units has been postulated
as early as 1932 by Zachariasen to describe the microscopic structure
of a-SiO$_2$ \cite{zach}. Apart from the construction of random
networks -- either by hand or by computer -- it is today also possible
to study relatively large models of a-SiO$_2$ which have been generated
by means of a molecular dynamics computer simulation in which the
sample was quenched with a cooling rate that is small compared to the
inverse of the time scale characteristic for atomic motion in fluids at
elevated temperatures \cite{voll1,voll2}.

Whereas there are numerous calculations of the electronic structure of
the crystalline phases of SiO$_2$, including recent self-consistent
computations within the local density approximation \cite{bingg,xu},
information on amorphous SiO$_2$ is sparse. Early work -- usually based
on local molecular orbital schemes 
-- has, e.g., been reviewed by Griscom
\cite{griscom}.  Laughin {\it et al.} have presented a tight-binding
study of a-SiO$_2$ on a Bethe lattice and compared it to
$\alpha$-quartz \cite{laughin}.  Gupta \cite{gupta} has used a
tight-binding approach to compare the electronic structure of
$\alpha$-quartz to a continuous random network (CRN) geometry
constructed by Dean and Bell \cite{dean} containing 614 atoms and a
structure generated by a molecular dynamics (MD) simulation of rapid
quenching, containing 648 atoms \cite{doan}.  Recently, Ordegon and
Yndurain have presented computations on both crystalline \cite{orde}
and amorphous \cite{yndu} modifications of SiO$_2$.  In these
calculations, matrix elements obtained from Hartree-Fock calculations
for small clusters have been transferred to a Bethe lattice, the density
of states has been obtained within a Green's function approach.
Compositional disorder was handled within an effective medium theory.

In the present work, we address the question of the electronic
structure of models of amorphous SiO$_2$ generated by molecular
dynamics simulations. The electronic structure is described by a simple
tight-binding Hamiltonian, including the crucial component of a
fluctuating Coulombic site energy, which to our knowledge has been
neglected in all previous approaches. We focus on the computation of
this fluctuation in the potential energy as a function of the cooling
rate with which the geometry was produced and its influence on the
density of states, the localization properties of eigenstates around
the former band gap and an approximate computation of the onset of the
optical absorption spectrum.

The rest of this article is organized as follows:  In the next section
we briefly outline the generation of the atomic configurations and give
a detailed description of the tight-binding electronic structure
calculation and its subsequent analysis. In the third section, results
are presented and discussed in comparison with experiments.
Conclusions are gived in the last section.


\subsection*{2. Methods}
\label{sec2}

The configurations of the ions used as input for the present electronic
structure calculations are the ones generated recently by Vollmayr {\it
el al.} in a molecular dynamics simulation of 
SiO$_2$~\cite{voll1,voll2}. 
The potential used was the one proposed by van Beest {\it et al.}
(BKS)~\cite{beest}, which has been demonstrated to give a good
description of the various crystalline phases of silica~\cite{kob29}.
The BKS potential contains only two-body interactions and reads
\begin{equation}
V(r_{ij}) = \frac{e^2}{4\pi \epsilon_0 } \frac{z_i z_j}{r_{ij}}
          + A_{ij} \exp(-B_{ij}r_{ij}) - \frac{C_{ij}}{r_{ij}^6}
\qquad.
\label{eq2.1}
\end{equation}
The values of the different parameters can be found in
Refs.~\cite{voll2,beest}.  For the computation of the electronic
structure presented here it is important to note that partial charges
of $z_+$=$z_{Si}$=2.4 and $z_-$=$z_O=-1.2$ have been used. Furthermore
we note that in the work of Vollmayr {\it et al.}, as well as in the
present work, the non-Coulombic part of the potential was truncated and
shifted at a distance of 5.5\AA. This truncation has the effect that
the density of the amorphous configurations is in agreement with the
experimental value of silica glass.

The goal of the work of Vollmayr {\it et al.} was to investigate how
the structural properties of silica depend on the rate with which the
sample was quenched, at constant pressure, from the liquid state at
high temperatures to an amorphous state at low temperatures. For this
the system was equilibrated at a high temperature $T_i=7000$ K and
subsequently coupled to a stochastic heat bath with temperature
$T_b(t)$. The temperature of this heat bath was decreased linearly in
time, i.e.  $T_b(t)=T_i-\gamma t$, where $\gamma$ is the cooling rate.
After $T_b$ reached 0 K, the so obtained configurations were relaxed
with a steepest descent method and subsequently analyzed. By varying
the cooling rate over more than two decades
($\gamma=$
1.1$\cdot$10$^{15}$ K/s,
5.7$\cdot$10$^{14}$ K/s,
2.8$\cdot$10$^{14}$ K/s,
1.4$\cdot$10$^{14}$ K/s,
7.1$\cdot$10$^{13}$ K/s,
3.6$\cdot$10$^{13}$ K/s,
1.8$\cdot$10$^{13}$ K/s,
8.9$\cdot$10$^{12}$ K/s and
4.4$\cdot$10$^{12}$ K/s)
it was possible to investigate how the properties of the glass, such as
various bond lengths or the distribution of the bonding angles, depend
on the cooling rate. In order to improve the statistics of the results
an average over ten independent runs was made for each value of
$\gamma$. At the beginning of the next section we will give a brief
summary of those results in order to facilitate the interpretation of
the results of the present electronic structure calculation. 
Details of the cooling rate simulation can be found in
Ref.~\cite{voll2}.

The electronic structure is described by a tight-binding
Hamiltonian
\begin{equation}
H = \sum_{ia} \epsilon_{ia} c^\dagger_{ia} c_{ia}
  + \sum_{i\ne j,ab} t_{ijab} c^\dagger_{ia} c_{jb}
\label{eq2.2}
\end{equation}
with creation/annihilation operators acting upon atomic orbitals
indexed $a,b$ centered at atoms $i,j$. A basis set of one s and three
p orbitals has been used for both silicon and oxygen atoms.
We use the diagonal and off-diagonal parameters $\epsilon^0$
given by Robertson
\cite{robert}: the differences between the diagonal energies
$\epsilon^0_{Si,s}$=5.1 eV,
$\epsilon^0_{Si,p}$=11.1 eV,
$\epsilon^0_{O,s}=-16.0$ eV and
$\epsilon^0_{O,p}=-1.08$ eV almost coincide with those
of the valence orbital
ionization potentials \cite{voip}.
Hopping matrix elements $t_{ijab}$ exist
between neighbor Si and O atoms. In addition, hopping
between two oxygen atoms bonded to the same Si atom is allowed. 
For Si-O bonds, matrix elements are generated using 
$V_{ss\sigma}$=--3.0 eV,
$V_{sp\sigma}$=5.2 eV,
$V_{pp\sigma}$=6.0 eV and
$V_{pp\pi}$=--1.0 eV; the only nonzero hopping
matrix elements for O-O bonds are
$V_{pp\sigma}$=0.56 eV and
$V_{pp\pi}$=--0.13 eV. Matrix elements between atomic orbitals
depend on the direction cosines and are modified according to the
Slater-Koster rules \cite{slater}.
Neighborhoods are defined by two cutoff radii, we use $r_1$=2.2 \AA \
for nearest and $r_2$=3.0 \AA \ for next-nearest neighbors, in
accord with the definition of neighborhoods in the analysis of
the microscopic structure of the same geometries \cite{voll2}. 

As will be elaborated below, in disordered systems with a strong ionic
character, the Madelung potential -- and thus the diagonal of the
Hamiltonian matrix -- depends on the center-of-mass coordinates due to
the underlying disorder. As we believe the average on-site
energies to be well represented by the parametrization scheme
described above, we consider fluctuations around these averages via
\begin{equation}
\epsilon_{ia}=\epsilon^0_{ia}+\langle V_M \rangle_i
- \frac{e^2}{4\pi \epsilon_0} \sum_{j \ne i} \frac{z_j}{r_{ij}}\quad .
\label{eq2.3}
\end{equation}
Here $V_M$ is the Madelung potential, the last term on the right hand
side.  We would like to note that we define the Madelung potential
entering the Hamiltonian~(\ref{eq2.2})  using an electronic test charge
placed at site $i$. The resulting large eigenvalue/eigenvector problems
of the size of 4008 basis functions have been solved using
storage-efficient Lanczos algorithms \cite{lanc}.

The localized or extended nature of eigenstates within the
former crystalline band gap decides the size of the mobility gap.
As a measure of electron localization, we use the modified inverse
participation ratio (MIPR). For normalized eigenstates
$\vert \alpha \rangle $, it is given by
\begin{equation}
{\mbox{MIPR}} = n^{1/2} \sum_i a_{i\alpha}^4, 
\label{eq2.4}
\end{equation}
where $a_{i\alpha}$ are the expansion coefficients of the eigenstates
$\vert \alpha \rangle = \sum_i a_{i\alpha } \vert i \rangle $.  The
MIPR is related to the inverse participation ratio, from which it
differs by the factor of the square root of the number of particles.
The participation ratio, $n^{1/2}/$MIPR, is a rough measure of the
number of atomic orbitals over which an eigenstate is extended.  The
scale invariance of the MIPR at the point of the transition from
localized to extended eigenfunctions has been attested by Chang et al.
\cite{chang}. For a variety of systems, an MIPR value close to unity
separates localized from extended states and thus marks the mobility
edge \cite{mobemp}.

Being interested in the contribution of different types of atomic
orbitals or of defects to eigenstates, we have performed a
population analysis. The Mulliken charge order
\begin{equation}
q_\alpha = \sum_i {\mbox{$^{'}$}} \langle \alpha \vert i \rangle
\langle i \vert \alpha \rangle = \sum_i {\mbox{$^{'}$}} a_{i\alpha}^2
\label{eq2.5}
\end{equation}
serves as a measure of the participation of a specific type
of orbitals -- hence the restriction of the sum indicated by a
prime -- in an eigenfunction $\vert \alpha \rangle$
\cite{mull}. In the
thermodynamic limit, the product of the charge order and
the density of the states equals the partial density of states (PDOS).

\subsection*{3. Results and discussion}
\label{sec3}

\subsubsection*{3.1 Microscopic structure}
\label{sec3.1}

In this subsection we briefly summarize some of the results of the work
of Vollmayr {\it et al.}~\cite{voll1,voll2} in order to facilitate the
understanding of the results of the present electronic structure
calculations.

As already mentioned in Sec. 2, the goal of that work was to
investigate how various properties of the glass depend on the cooling
rate $\gamma$ with which the glass was quenched from the liquid state.
Regarding the structural properties it was found that the peaks in the
various radial distribution functions, or in the partial structure
factors, become more pronounced with decreasing $\gamma$. In particular
it was shown that the so-called first sharp diffraction peak in the
structure factor increases significantly, which shows that the
arrangement of the ions become more ordered not only locally, i.e. on
the length scale of one  Si-O bond, but also on the intermediate range
length scale, i.e. on the length scale of a few tetrahedra. This
observation was corroborated by the fact that also the distribution of
the various bond angles, such as the intra-tetrahedral angle O-Si-O, or
the inter-tetrahedral angles Si-O-Si and Si-Si-Si showed significantly
more pronounced peaks with decreasing cooling rate. Furthermore it was
found that the distribution of the rings of a given size depends
strongly on the cooling rate in that with decreasing $\gamma$ the
frequency of very short rings ($n\leq 3$), as well as the ones of the
long ones ($n \geq 8$), decreases. Here $n$ is the number of Si-O pairs
in a ring. On the other side the frequency of rings with length $n=5$
and 6 increases with decreasing cooling rate, which can be understood
by remembering that at normal pressures the crystalline phase of silica,
$\beta$-crystobalite, purely consists of six-membered rings.
Thus it is reasonable
to assume that with decreasing cooling rate the {\it local} structure
of the glass is quite similar to the one of $\beta$-crystobalite.

\subsubsection*{3.2 Madelung potential fluctuations}
\label{sec3.2}

Potential energy fluctuations play a crucial role in the understanding
of electronic properties of ionic liquids.  Due to the disorder in the
positions of the ionic centers of mass, the Madelung potential $V_M$
is not a constant as in an ideal
crystalline system, but a function of the ionic coordinates. This
concept has been introduced by Logan and Siringo \cite{logan1} to ionic
fluids. These authors have used the mean spherical approximation (MSA)
to compute the fluctuations $\Delta V_M$ of $V_M$ for the restricted
primitive model at high density, as relevant for molten CsAu. At the
melting point of simple ionic systems, $\Delta V_M$ is of the order of
one electron Volt; linear graphic theories of liquids like the MSA
predict a Gaussian distribution of potential
energies \cite{logan1}. The impact of a random Madelung potential upon
the electronic structure of fluids has been studied numerically for
liquid Cs$_x[$CsAu$]_{1-x}$ \cite{tkcsau}, molten alkali halides like
KCl \cite{tkmx}, alkali halide - alkali metal solutions
K$_x[$KCl$]_{1-x}$ \cite{tkmmx} and liquid silver chalcogenides
\cite{tkag2x}.

Having generated SiO$_2$ geometries by molecular dynamics simulations
we are able to address the question whether these fluctuations persist
once a fluid is rapidly quenched into the amorphous state.  Using an
electronic test charge and full formal charges $z_+$=4 and $z_-=-2$, we
have computed the anionic and cationic Madelung potentials and their
fluctuations. In Fig.~1 we show $\Delta V_M$ as a function of
the cooling rate $\gamma $. We recognize that the Madelung potential
fluctuations are increasing with increasing $\gamma$ and that they are
on the order of $\sim$1.75 eV for the fastest and $\sim$1 eV for the
slowest cooling rate studied.  In general, fluctuations in the anionic
Madelung potential are stronger than those in their cationic
counterpart, a tendency that seems to increase with increasing cooling
rate.  Compared to fluids at elevated temperatures \cite{tkun},
fluctuations in $\Delta V_M$ for quenched SiO$_2$ from one realization
to the other are surprisingly large, the corresponding RMS deviation is
of the order of 0.15 eV, as depicted by the error bar Fig.~1 for the
cationic $\Delta V_M$ with the largest value of $\gamma $. Even with
the large amount of data used in this study we see no way of performing
a reliable extrapolation to a cooling rate which is typical in a real
experiment.  Any reduction of the formal
charge $z_1$ to an effective charge $z_2$ will reduce the potential
energy fluctuations accordingly, 
since the RMS deviations obey the relation
$\Delta V_M(z_1)/\Delta V_M(z_2) = z_1/z_2$. Thus any possible reduction
of the Coulombic site energy fluctuations by a slower quenching
procedure can be partly compensated by using the full formal charge --
the customary procedure when one computes Madelung potentials --
instead of the simulation charge as frequently done in this work.

\subsubsection*{3.3 Density of states}
\label{sec3.3}

To discuss the influence of topological disorder, bond angle
fluctuations and Made\-lung potential fluctuations upon the density of
states, we use the crystalline DOS -- as computed by Robertson
\cite{robert} using the same set of parameters as in that work -- which
is shown in Fig.~2a as a reference.  The tight-binding parameters are
chosen to make the upper edge of the valence band the zero of energy.
The narrow band around --18 eV is dominated by oxygen 2s orbitals, a
structured valence band of width $\sim $13 eV is separated from the
conduction band by a band gap of $\sim $9 eV, in good agreement with
the experimental value of the conductivity gap of $\sim $9.4 eV in
$\alpha $-quartz \cite{cond}. States at the top of the valence band
display a strong contribution of oxygen 2p orbitals and exhibit a
nonbonding character.  Details of the models of a-SiO$_2$ used for
electronic structure computations in this work are also listed in table
1.

As mentioned in Sec. 3.1, in our system the disorder manifests
itself in the distribution of bond lengths, bond angles and in the
possible deviation of the size of (SiO)$_n$ rings from those observed
in crystalline systems, including the presence of odd-membered rings.
In a computer experiment, it is straightforward to study the influence
of the disorder on the density of state even if the fluctuations in the
Madelung potential are absent and in Fig.~2 we present the
result of such a calculation.

For the smallest cooling rate ($\gamma $=4.4$\cdot$10$^{12}$ K/s), the
resulting density of states is displayed in Fig.~2b. In contrast to
the alkali halides \cite{tkmx}, the main features of the bands are
remarkably persistent with respect to the introduction of disorder.
Whereas the fine structure of all bands is smoothed out considerably,
usually attributed to the breakdown of topological disorder 
\cite{ring}, a significant broadening of the
bands can only be found at the bottom of the valence band, the bottom
of the oxygen 2s band and at the top of the conduction band. For the
energies at which changes in the density of states have the most
profound physical influence
 -- in the gap around the Fermi level -- no tailing can be observed
within the resolution of the DOS as presented here (0.25 eV). A small
number of eigenstates (two out of 40080 for $\gamma
$=4.4$\cdot$10$^{12}$ K/s and ten for $\gamma $=1.1$\cdot$10$^{15}$
K/s) have been found at energies around 3-4 eV. Inspecting the
corresponding eigenvectors, these states can be unambiguously
attributed to the highest occupied orbitals of four-membered (SiO)$_2$
rings.

This situation changes considerably once Madelung potential
fluctuations are introduced (Fig.~2c). Using the full formal charges
$z_{Si}$=+4 and $z_O$=--2 on the ions, we observe a considerable
tailing of both the upper valence band edge and the lower conduction
band edge into the former band gap. The amount of band tailing
evidently depends on the cooling rate and thus on the thermal history
of the model system: the solid line corresponds to a cooling rate of
$\gamma $=4.4$\cdot$10$^{12}$ K/s, as in Fig.~2b, the dotted line to a
$\gamma $ value of 1.1$\cdot$10$^{15}$ K/s.  The question of the
localized or extended nature of states in the gap -- crucial for the
existence and the size of a mobility gap -- will be addressed in the
next section.

\subsubsection*{3.4 Localization properties and the mobility gap}
\label{sec3.4}

As stated above, in this work the modified inverse participation ratio
[MIPR, Eq.~(\ref{eq2.4})] serves as a measure to determine whether the
nature of eigenstates around the Fermi energy is localized or
extended.  With a system size restricted to 1002 atoms, we did not
study the dependence of localization measures upon the system size, but
took a MIPR value larger than unity as an indicator of localized
states, and a MIPR value smaller than unity characterizing extended
states. As will be shown below, the crossover between localized and
extended states is sharp for the large systems studied here, so the
error made for the size of the mobility gap resulting from small
deviations of the critical MIPR from the value of unity should be small.

The MIPR as a function of energy is presented in Fig.~3. With Madelung
potential fluctuations switched off ($\gamma $=4.4$\cdot$10$^{12}$ K/s,
$\nabla $) we observe the following localization behavior: At the
bottom of the conduction band, only extended states exist. This is a
clear numerical verification of the hypothesis that neither bond angle
nor topological disorder are sufficient to induce localization within
an energy interval larger than some tenths of an electron volt for
eigenstates with a strongly bonding character \cite{topol}. At the top
of the nonbonding oxygen band, a crossover from extended to localized
states can be observed. It is, however, important to note that only two
out of 40080 eigenstates lie in the energy interval at the top of the
valence band. Thus we see that in the absence of Madelung potential
fluctuations, localized eigenstates dominate an energy interval of some
tenths of an electron Volt at the top of the valence band, that their
total number is, however, negligibly small. The mobility gap and the
energy interval containing no density of states almost coincide.

In the presence of Madelung potential fluctuations, the situation is
changed dramatically. For the same cooling rate of $\gamma
$=4.4$\cdot$10$^{12}$ K/s ($\times $) as before, the MIPR crosses its
critical value of unity both at the top of the valence and at the
bottom of the conduction band, a broad range of localized eigenstates
appear in the former band gap. The size of the mobility gap is
approximated as $\sim $9.1 eV. The most localized states deep in the gap
are extended over little more than an oxygen atom, thus are close to
the maximum degree of localization. If the amount of disorder
present in the structure is increased, by increasing the cooling rate to
1.1$\cdot$10$^{15}$ K/s ($\circ $,solid line), no qualitative change of
the energy dependence of the MIPR is observed, albeit the statistics
has improved, due to the larger number of eigenstates in the former gap
(cf.  Fig.~2c). The magnitude of the mobility gap increases slightly.

To summarize, Madelung potential fluctuations provide the only
mechanism for the creation of eigenstates deep in the former gap within
the model systems studied here. The size of the mobility gap dominated
by localized states is of the order of 9 eV, close to the experimental
crystalline conductivity gap of 9.4 eV ($\alpha$-quartz), the onset of
photoconductivity in a-SiO$_2$ with an upper boundary of 9.0 eV
\cite{photo}, and the photoinjection threshold of 8.9$\pm$0.2 eV, which
is not sensitive to selection rules \cite{photo}.

\subsubsection*{3.5 Band tails and the optical gap}
\label{sec3.5}

The experimentally observed photoconductivity of a-SiO$_2$ of 9.0 eV is
thus consistent with two types of the density states presented in this
work:  i) in the absence of Madelung potential fluctuations (Fig.~2b,
Fig.~3, $\nabla$ symbol) the mobility gap almost coincides with the
band gap, topological disorder does neither lead to a significant band
tailing nor to localization at the band edges; or ii) in the presence
of Madelung potential fluctuations, localized eigenstates exist deep
within the former band gap (Fig.~2c, Fig.~3, $\circ$ and $\times $
symbols), and the onset of the photoconductivity originates from a large
mobility gap of $\sim$ 9 eV.  In this section, we will investigate
whether one of the two pictures has to be abandoned if the optical
absorption properties of a-Si are considered.

The imaginary part of the dielectric function, $\epsilon_2$, governing
the optical absorption, can be approximated by the joint
density of states, JDOS, divided by the excitation frequency
\cite{jdos}:
\begin{equation}
\epsilon_2 (\omega) \propto \omega^{-1} 
\int_{-\infty}^{E_F} {\mbox{DOS}}_{VB}(E)
{\mbox{DOS}}_{CB}(E+\hbar \omega) dE, 
\label{eq3.1}
\end{equation}
where ${\mbox{DOS}}_{VB}$ and ${\mbox{DOS}}_{CB}$ 
denote the valence and the
conduction band density of states.
This approach neglects variations of the dipole transition matrix
element with energy, an approximation considered to be particularly
severe in the case of SiO$_2$ \cite{dipole}.  In the present work,
however, our primary concern is the shape of the absorption edge
originating from eigenstates that can be located within the former
crystalline band gap. In this energy range, the JDOS will vary over
several orders of magnitude. Variations in the dipole transition matrix
element will be comparatively small since
due to the random nature of the model systems, there exist no
transitions that is strictly forbidden by symmetry. The first optical
band in crystalline SiO$_2$ is usually attributed to an exciton at 10.4
eV \cite{exciton}. The experimental determination of the indirect band
gap is thus restricted to the measurement of the photoconductivity. The
first clear 
peak of the optical absorption spectrum can be observed at 11.7
eV \cite{harropt}. 

The geometries obtained from MD simulations still contain a small
number of defects, which may spoil the comparison of the computed
absorption to experiments with those samples of amorphous SiO$_2$ where
defects are believed to be removed by the addition of water
\cite{apl}.
As the band tailing is
primarily caused by Madelung potential fluctuations, we simply
superimpose potential energy fluctuations obeying a Gaussian
distribution on a model system with the geometry of
$\alpha$-crystobalite. To compute a property originating from an
integration over the DOS, 25 realizations of systems containing 638
atoms turn out to be sufficient. The resulting optical absorption is
presented in Fig.~4. Absorption edges associated with systems
exhibiting Madelung potential fluctuations show a significant tail, the
onset and slope of which is determined by the RMS deviation of the
Coulombic potential energy.  With $\Delta V_M^+ = \Delta V_M^-$=0.6 eV,
corresponding to the use of the charges used within the MD simulations
and a conservative estimate of the strength of potential energy
fluctuations (see Sec.~3.2), the onset of the absorption can
be located around 7~ eV ($\nabla$), full formal charges give rise to an
onset at 5.5~eV ($+$), and a reduction of the charge to half of the
formal charge shifts the onset of the spectrum at 8 eV ($\times $).  As
all spectra show an exponential tail, there is no clear termination of
the absorption spectra, the overall trend, however, is evident from
Fig.~4. Removing the aforementioned defect states associated with the
presence of four-membered rings, a band edge in the absence of Madelung
potential fluctuations is also presented in Fig.~4 ($\circ $, $\gamma
$=4.4$\cdot$10$^{12}$ K/s).  An energy as high as 9 eV is required to
induce a small absorption for a system exhibiting only bond angle and
topological disorder.

The experimentally observed onset of the absorption spectrum crucially
depends on the water content of the sample \cite{apl}:  glasses
characterized by a small concentration of water shows an absorption
peak at 7.6 eV preceding a continuous absorption spectrum. The
intensity of this maximum depends on the sample studied. No such peak
is found for glasses characterized by a comparatively high
concentration of water; again depending on the type of silica, the
onset of the absorption spectrum can be located around 7.0$\pm$0.2 eV
\cite{apl}. The role of H$_2$O has been interpreted as saturating
defects like dangling bonds and thus removing the corresponding
eigenstates from the band gap \cite{photo}. This interpretation
encourages us to compare the absorption edges obtained from samples
with a high concentration of water -- moving defect states into the
valence or conduction band by the formation of a strong chemical bond
-- to our model free of defects, which is governed by fluctuations of
the Coulombic site energies.  It is evident from Fig.~4 that Madelung
potential fluctuations of 0.6~eV -- as induced by the charges used in
the pair potential Eq.~(\ref{eq2.1}) -- are sufficient to shift the
absorption edge to the red by two electron Volts, now coinciding with
the experimental value.

It has to be noted that the precise form of the band edges
computed via iterative schemes like the recursion method, depends
on the way the continued fraction is terminated, so we see no way
of getting around the laborious way of studying a large number of 
slowly quenched large systems or carefully equilibrated CRNs by exact
diagonalization. Naturally, the same argument holds for the 
computation of localization properties of eigenstates within the
pseudo-gap.

\subsubsection*{3.6 The role of defects}
\label{sec3.6}

As stated in Sec. 3.3, only an extremely small fraction of defects
associated with the presence of four-membered (SiO)$_2$ rings can be
located within the former band gap, if the fluctuations of the Madelung
potential are absent. To our knowledge, there is no experimental
evidence either of these states or of small rings in amorphous silica.
There is no connection between these states and any type of over- or
under-coordinated silicon or oxygen atom. We believe that these rings
will be suppressed by further decreasing the cooling rate and in
Ref.~\cite{voll2} evidence was given for this.

Once Madelung potential fluctuations are introduced, the fraction of
gap states localized on four-membered rings can be neglected compared
to the strong tailing of states at the band edges. Computing the defect
charge order -- Eq.~(5) with the sum restricted to
orbitals localized on under- or over-coordinated atoms -- 
we notice that these defect basis
functions now make a major contribution to localized eigenstates: the
total charge order of defects peaks at a value close to 1/2 at an
energy of $\sim$6 eV, as shown in Fig.~5. As also evident from this
figure, under-coordinated oxygen atoms dominate the defect density of
states.  It is remarkable that Madelung potential fluctuations do not
simply broaden a defect band within the gap, but seem to be essential
for the presence of defect states. A possible explanation for this
behavior may be given by the strong localization of eigenstates in a
random potential, reducing the coupling of dangling bond orbitals to
the rest of the system and thus preventing these states from being
split into high energy CB and low energy VB states by forming strong
covalent bonds.

Note that these computations were performed 
for systems characterized by a
cooling rate of 1.1$\cdot$10$^{15}$ K/s, the small number of
eigenstates deep within the gap does not permit us to make any
statement based on proper statistics for the slowest cooling rate of
4.4$\cdot$10$^{12}$ K/s.

\subsection*{4. Conclusions}
\label{sec4}

We have presented a numerical study of the electronic structure of
models of amorphous silica. The geometries used have been generated
with a molecular dynamics simulation in which the system was quenched
with different cooling rates, thus had a variable amount of disorder.
The electronic structure has been described by a tight-binding
Hamiltonian incorporating fluctuations arising from the disordered
ionic center-of-mass positions, leading to variations in the Coulombic
site energies with an RMS deviation of the order of 0.6 eV for the
charges used in the MD simulation and the smallest cooling rates
applied.

The influence of disorder upon the density of states, localization
properties and the size of the optical gap are studied and discussed.
Madelung potential fluctuations lead to the formation of a mobility gap
of the order of 9 eV around the Fermi level, which turns out to be
insensitive to any physically reasonable variation of the amount of
disorder present in the model. The size of this energy interval
containing only localized states not contributing to electronic
transport is in accord with the experimental value of the
photoconductivity threshold (9 eV) and with photoinjection experiments
(8.9$\pm$0.2 eV) \cite{photo}.

In the absence of Madelung potential fluctuations, band tailing is
suppressed, and the amount of localized states within the former gap is
vanishingly small. This behavior would imply an optical gap of the
order of 9 eV, whereas \mbox{a-SiO$_2$} optimized for transparency in
the UV region exhibits an onset of the optical absorption around 7 eV
\cite{apl}. A defect-free model incorporating only site disorder -- a
simple model to mimic the removal of defect states from the gap by
increasing the content of water -- does, however, reproduce the
numerical value of the gap, again using the charges applied within the
MD simulations.

To conclude, fluctuations in the Madelung potential -- a key element in
the theory of the electronic structure of ionic fluids \cite{logan1} --
persist in models of quenched silica at least for cooling rates down to
the order of 10$^{12}$ K/s.  This concept arising from liquid-state
theory turns out to be crucial for the understanding of the nature of
electronic eigenstates within the gap.  It provides a possible
mechanism to induce the strong mismatch of the mobility gap and the
optical gap of 2 eV, as observed in a-SiO$_2$.

\subsubsection*{Acknowledgments}

It is a pleasure to thank U. Beck, W. Freyland and D. Nattland for
fruitful discussions and A. K\"onig for inspiration.  Financial support
by the Deutsche Forschungsgemeinschaft (grants Ko 1384/2-2 and through
the SFB 262) is gratefully acknowledged. K.V. thanks Schott-Glaswerke
Mainz for financial support through the Schott-Glaswerke-Fonds. Part
of this work has been performed on computer facilities of the
Regionales Rechenzentrum Kaiserslautern.


\newpage
\subsection*{Table 1}

Models of quenched SiO$_2$ used for electronic
structure calculations. All amorphous geometries originating
from MD simulations contain 1002 atoms (10 realizations),
lattice geometries contain 638 atoms (25 realizations).

\begin{tabular}{|c|l|c|c|l|c|c|c|}
\hline
Model & Model & $\Delta V_M^+ $ & $\Delta V_M^-$
& $\gamma $ & fig.2 & fig.3 & fig.4 \\
index & geometry & (eV)            & (eV)
& (K/s) &       &    &  \\
\hline
1 & amorphous & 0.89 & 0.93 & 4.4$\cdot$10$^{12}$ & c 
& $\times $ & \\ 
2     & amorphous & 1.58 & 1.88 & 1.1$\cdot$10$^{15}$ 
& c & $\circ $  & \\ 
\hline
3     & amorphous & zero & zero & 4.4$\cdot$10$^{12}$ & b & 
$\nabla $
& $ \circ $\\ 
4     & amorphous & zero & zero & 1.1$\cdot$10$^{15}$ & & & \\
\hline
5     & lattice & 0.3 & 0.3 & --  & & & $\times $ \\
6     & lattice & 0.6 & 0.6 & --  & & & $\nabla $ \\
7     & lattice & 1.0 & 1.0 & --  & & & $+$ \\
\hline
\end{tabular}
\newpage
\subsection*{Figure captions}

{\bf Figure 1} Cationic ($\circ$) and anionic ($\times $) Madelung
potential fluctuations, $\Delta V_M$, as a function of the logarithm of
the cooling rate, $\gamma $. Solid and dotted lines are a guide to the
eye. 


{\bf Figure 2} Silica density of states for various degrees of
disorder: a) crystalline $\alpha$-quartz (after \cite{robert}), b)
amorphous silica, $\gamma$=4.4$\cdot$10$^{12}$ K/s, without Madelung
potential fluctuations, c) amorphous silica with Madelung potential
fluctuations (solid line: $\gamma $=4.4$\cdot$10$^{12}$ K/s, dotted
line:  $\gamma $=1.1$\cdot$10$^{15}$ K/s).


{\bf Figure 3}
Logarithm of the modified inverse participation ratio as a function of
energy.  $\nabla $: $\gamma $=4.4$\cdot$10$^{12}$ K/s, no Madelung
potential fluctuations; $\times $:  $\gamma $=4.4$\cdot$10$^{12}$ K/s,
including Madelung potential fluctuations; $\circ $: $\gamma
$=1.1$\cdot$10$^{15}$ K/s, including Madelung potential fluctuations
(solid line is a guide to the eye).


{\bf Figure 4} Logarithm of the joint density of states divided by
the excitation frequency as a function of energy:
$+$: lattice model, cationic charge $z_+$=4;
$\nabla $: lattice model, $z_+$=2.4;
$\times $: lattice model, $z_+$=1.2;
$\circ $: quenched SiO$_2$, 
$\gamma $=4.4$\cdot$10$^{12}$ K/s, no Madelung potential
fluctuations. Solid and dotted lines are a guide to the eye.


{\bf Figure 5} Dimensionless defect charge order as a function of
energy for $\gamma $=1.1$\cdot$10$^{15}$ K/s:  total ($\circ $, solid
line) and singly bonded oxygen charge order ($\times $, dotted line).
Lines are a guide to the eye;


\newpage
\begin{thebibliography}{99}
\bibitem{zach} W.H. Zachariasen, J. Am. Chem. Soc. {\bf 54}, 3841
   (1932).
\bibitem{voll1} K. Vollmayr and W. Kob, 
   Ber. Bunsenges. Phys.  Chemie {\bf 100}, 1399 (1996).
\bibitem{voll2} K. Vollmayr, W. Kob and K. Binder, Phys. Rev. B 
  {\bf 54}, 15808 (1996).
\bibitem{bingg} N. Binggeli, N. Troullier, J.C. Martins, and
   J.R. Chelikowsky, 
   Phys. Rev. B {\bf 44}, 4771 (1991) and references therein.
\bibitem{xu} Yong-nian Xu and W.Y. Ching,
   Phys. Rev. B {\bf 44}, 11049 (1991).
\bibitem{griscom} D.L. Griscom, J. Non-Cryst. Sol. {\bf 24}, 155
   (1977).
\bibitem{laughin} R.D. Laughin, J.D. Joannopoulos, and D.J. Chadi,
   Phys. Rev. B {\bf 20}, 5228 (1979).
\bibitem{gupta} R.P. Gupta,
   Phys. Rev. B {\bf 32}, 8278 (1985).
\bibitem{dean} R.J. Bell and P. Dean,
   Phil. Mag. {\bf 25}, 1381 (1972).
\bibitem{doan} 
   N.V. Doan, Phil. Mag. A {\bf 49}, 683 (1984).
\bibitem{orde} P. Ordejon and F. Yndurain,
   Phys. Rev. B {\bf 43}, 4552 (1991).
\bibitem{yndu}
F. Yndurain and P. Ordejon,
   Phil. Mag. B {\bf 70}, 535 (1994).
\bibitem{beest} B.W.H. van Beest, G.J. Kramer, and R.A. van Santen,
   Phys. Rev. Lett. {\bf 64}, 1955 (1990); a similar potential
   has been proposed by S. Tsuneyuki, M. Tsukuda, H. Aoki and
   Y. Matsui, Phys. Rev. Lett. {\bf 61}, 869 (1988).
\bibitem{kob29} J.S. Tse and D.D. Klug, 
   Phys. Rev. Lett. {\bf 67}, 3559 (1991);
   J. Chem. Phys. {\bf 95}, 9176 (1991);
   J.S. Tse, D.D. Klug and Y. LePage,
   Phys. Rev. B {\bf 46}, 5933 (1992);
   Phys. Rev. Lett. {\bf 69}, 3647 (1992);
   J.S. Tse, D.D. Klug and D.C. Allen,
   Phys. Rev. B {\bf 51}, 16392 (1995).
\bibitem{robert} J. Robertson, Adv. Phys. {\bf 32}, 361 (1983).
\bibitem{voip} F. Herman and S. Skillman, {\sl Atomic structure
   calculations} (Prentice Hall, Englewood Cliffs NY, 1963).
\bibitem{slater} J.C. Slater and G.F. Koster, 
   Phys. Rev. {\bf 94}, 1498 (1954).
\bibitem{lanc} C. Lanczos, J. Res. NBS B {\bf 45}, 225 (1950);
   J.K. Cullum and R. Willoughby, {\sl Lanczos algorithms for 
   large symmetric eigenvalue problems: I, theory; II, programs},
   Birkh\"auser, Boston 1985;
   Th. Koslowski and W. von Niessen, J. Comp. Chem. {\bf 14},
   769 (1993). 
\bibitem{chang} T.-M. Chang, J.D. Bauer, J.L. Skinner,
   J. Chem. Phys. {\bf 93}, 8973 (1990).
\bibitem{mobemp} Th. Koslowski, D.G. Rowan and D.E. Logan,
   Ber. Bunsenges. Phys.  Chemie {\bf 100}, 101 (1996);
   H. D\"ucker, Th. Koslowski, W. von Niessen, M.A. Tusch and
   D.E. Logan, J. Non-Cryst. Solids {\bf 207}, 32 (1996).
\bibitem{mull} Mulliken, R.S.; {\sl J. Chem. Phys} {\bf 1955},
23, 1833
\bibitem{logan1} D.E. Logan and F. Siringo,
  J. Phys. Cond. Matter {\bf 4}, 3695 (1992);
  F. Siringo and D.E. Logan, J. Phys. Cond. Matter {\bf 5}, 
  1841 (1993).
\bibitem{tkcsau} Th. Koslowski and D.E. Logan, 
   J. Phys. Chem. {\bf 98}, 9146 (1994).
\bibitem{tkmx} Th. Koslowski,
   Ber. Bunsenges. Phys.  Chemie {\bf 100}, 95 (1996).
\bibitem{tkmmx} Th. Koslowski, J. Chem. Phys., in press.
\bibitem{tkag2x} Th. Koslowski, 
  J. Phys. Cond. Matter {\bf 8}, 7031 (1996).
\bibitem{tkun} Th. Koslowski, unpublished, 1996.
\bibitem{cond} W. Feltz, {\sl Amorphous inorganic materials
   and glasses}, p.217 (VCH, Weinheim, 1993).
\bibitem{ring} D. Weaire and M.F. Thorpe, in F. Herman, 
  A.D. MacLean and R.K. Nesbet eds. {\sl Computational methods
  for large molecules and localized states in solids}, p.295
  (Plenum Press, New York, 1973).
\bibitem{topol} J. Singh, Phys. Rev. B {\bf 23}, 4156 (1981).
\bibitem{photo} T.H. diStefano and D.E. Eastman, 
   Sol. State. Comm. {\bf 9}, 2259 (1971).
\bibitem{jdos} See e.g. W.A. Harrison, {\sl Electronic structure and
the properties of solids}, 2nd ed., pp.103, Dover Publishing,
N.Y. 1989.
\bibitem{dipole} J.R. Chelikowsky and M.R. Schl\"uter,
   Phys. Rev. B {\bf 15}, 4020 (1977).
\bibitem{exciton} K. Platz\"oder, Phys. Stat. Sol. {\bf 29}, 63
   (1968).
\bibitem{harropt} G. Klein and H.O. Chun,
  Phys. Stat. Sol. b {\bf 49}, 167 (1972).
\bibitem{apl} I.P. Kammow, B.G. Bavgely and C.G. Olsen,
  Appl. Phys. Lett {\bf 32}, 98 (1978).
\end{thebibliography}
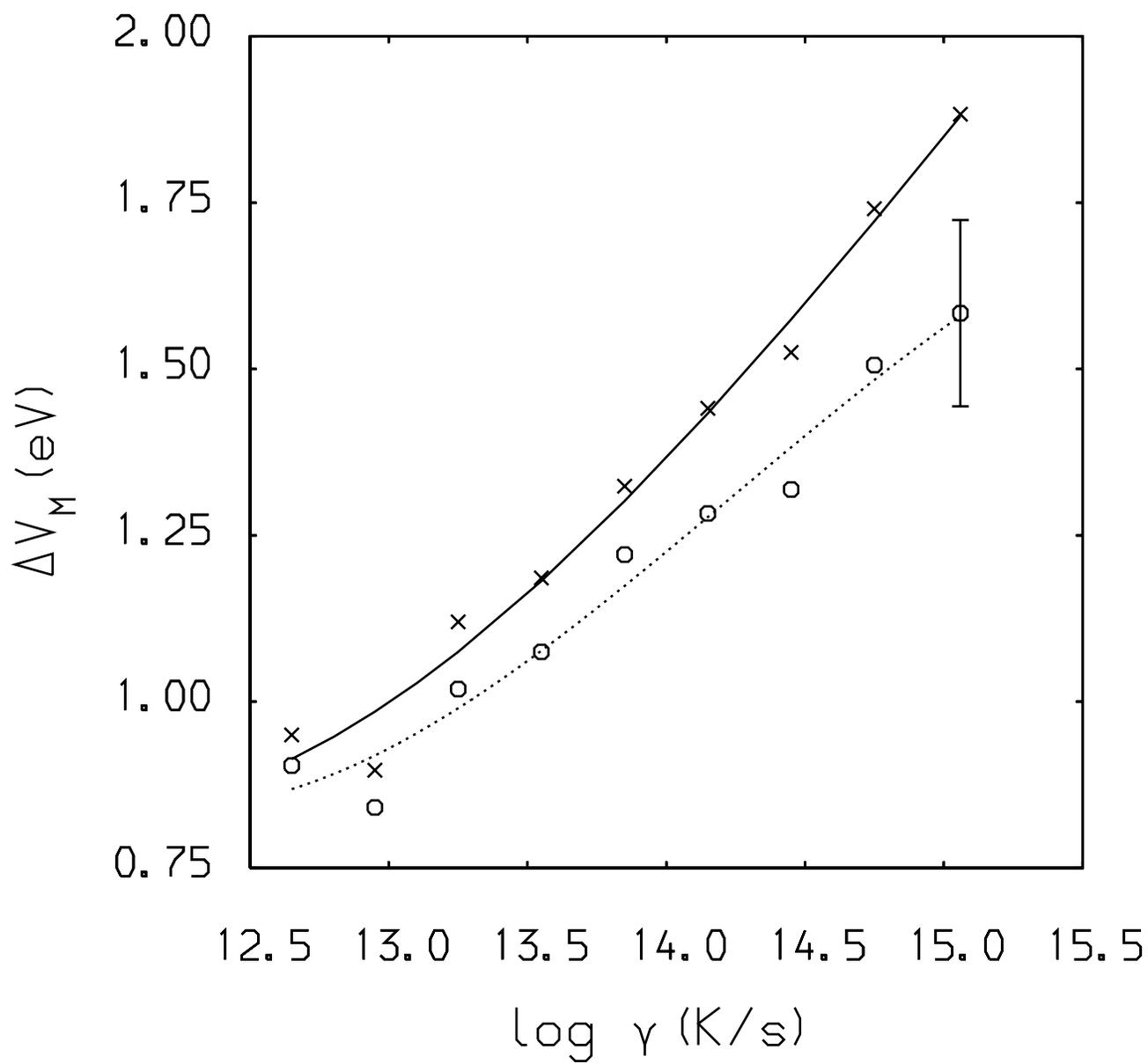
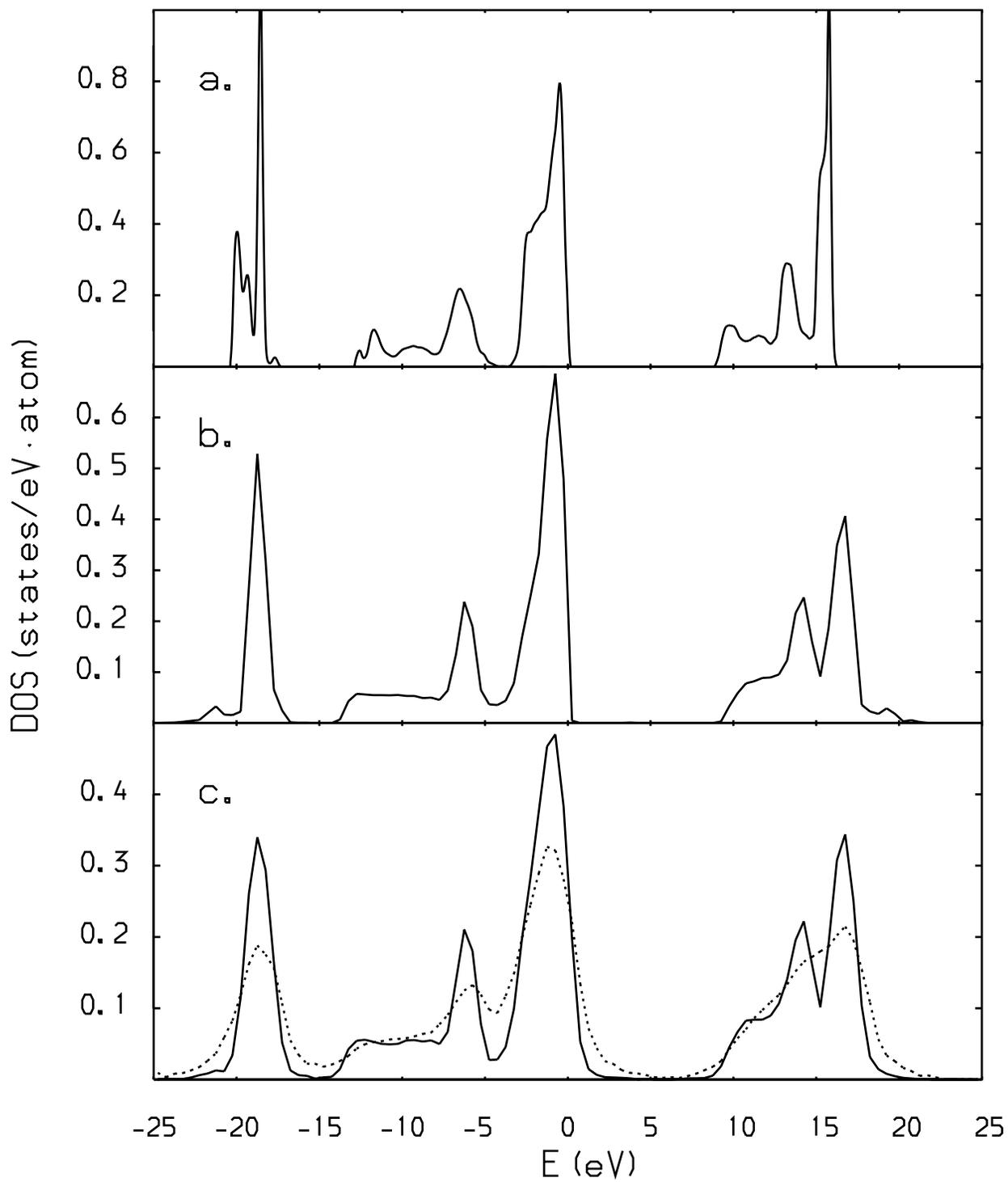
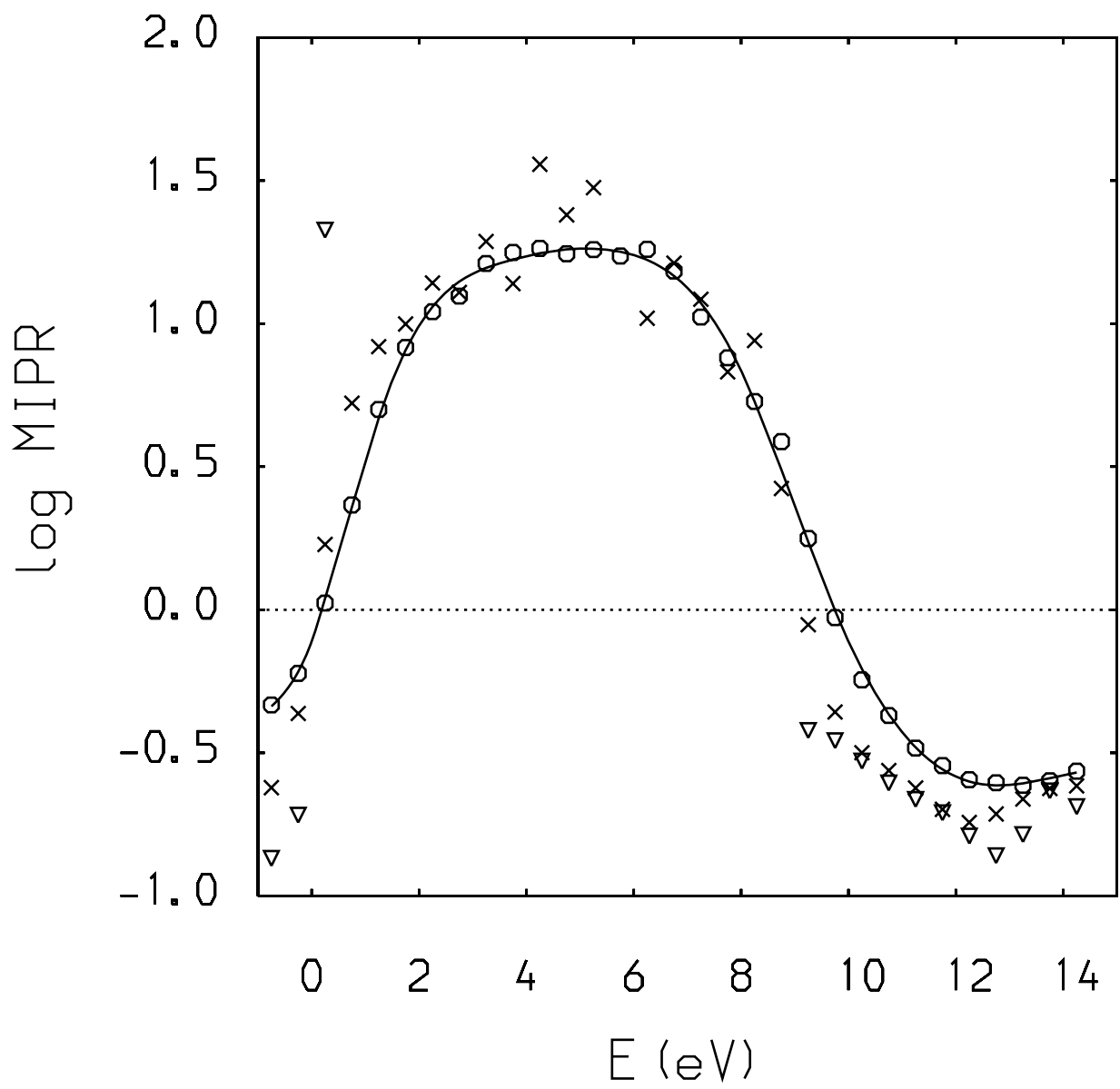
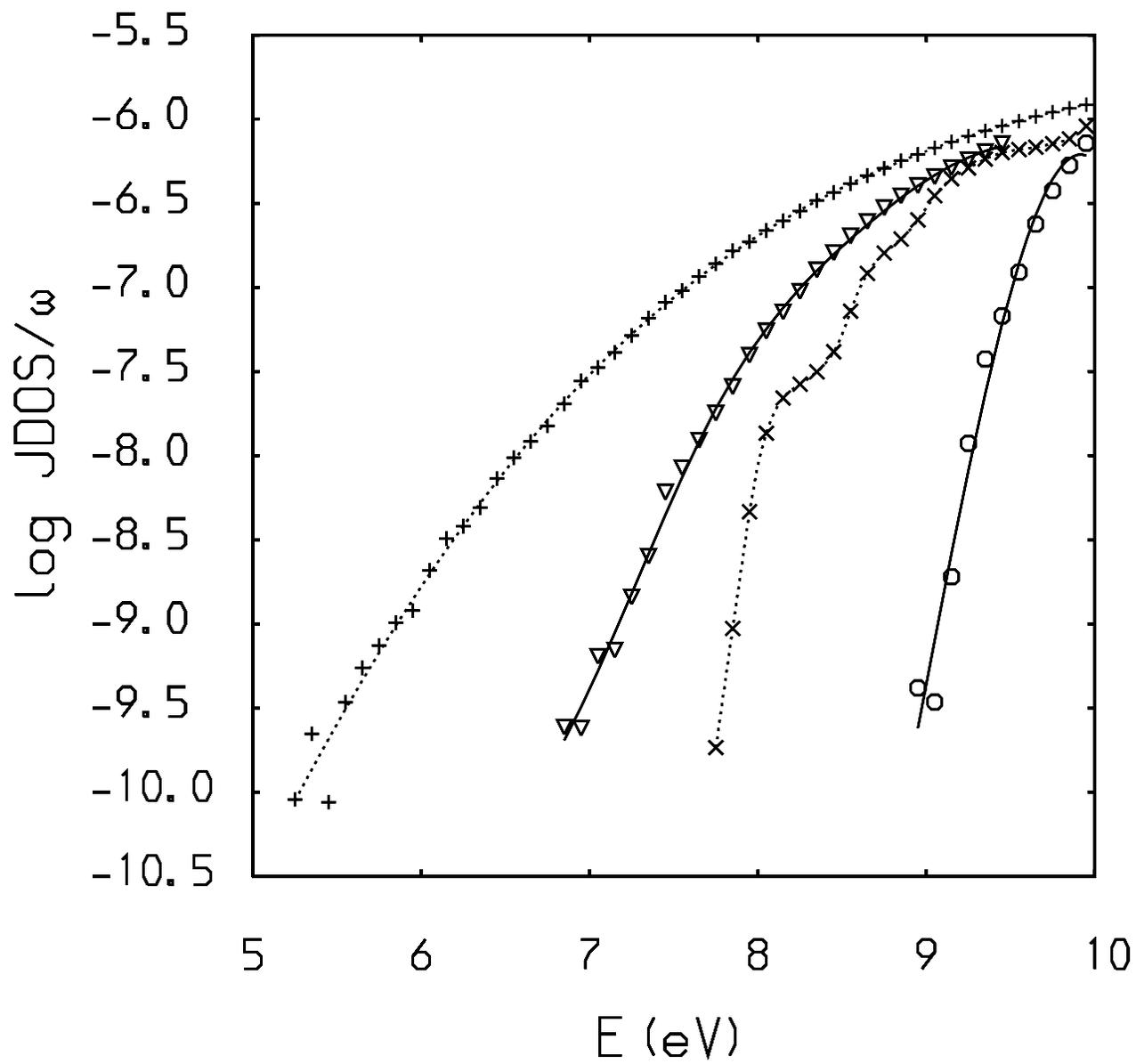
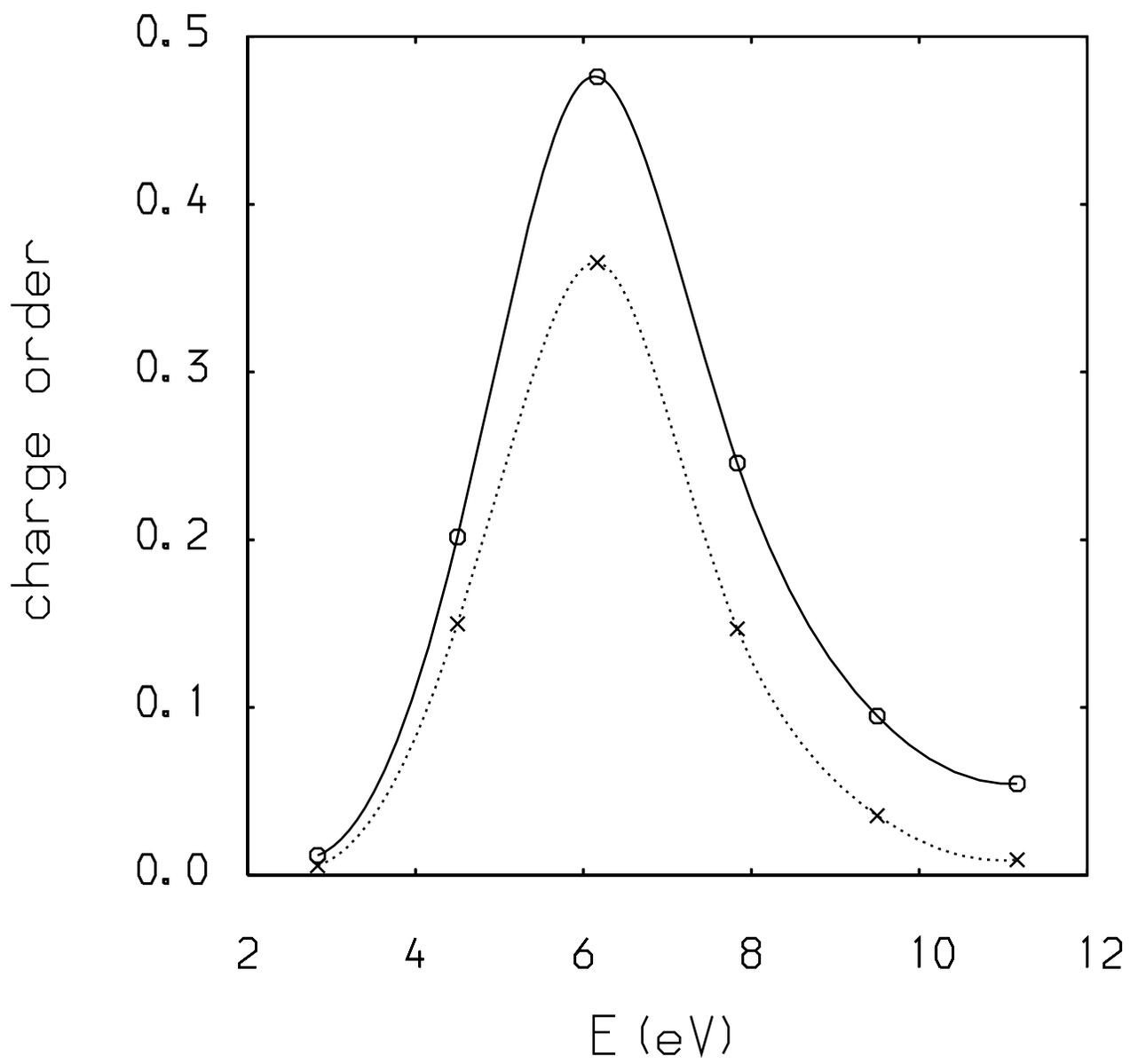
\end{document}